\documentclass[10pt,a4paper,twocolumn]{article}

\usepackage[utf8]{inputenc}
\usepackage[T1]{fontenc}
\usepackage{amsmath}
\usepackage{amsfonts}
\usepackage{amssymb}
\usepackage{graphicx}
\usepackage{hyperref}
\usepackage{natbib}
\usepackage{booktabs}
\usepackage{float}
\usepackage{geometry}
\usepackage{fancyhdr}
\usepackage{algorithm}
\usepackage{algpseudocode}
\usepackage{multicol}

\title{\textbf{Neural Network-Based Algorithmic Trading Systems: \\
Multi-Timeframe Analysis and High-Frequency Execution \\
in Cryptocurrency Markets}}

\author{
    \textit{Zhāng Wěi}\\
    \textit{zhang@polar-tensor.com}
}

\begin{document}

\twocolumn[
\begin{@twocolumnfalse}
\maketitle

\begin{abstract}
This paper explores neural network-based approaches for algorithmic trading in cryptocurrency markets. Our approach combines multi-timeframe trend analysis with high-frequency direction prediction networks, achieving positive risk-adjusted returns through statistical modeling and systematic market exploitation. The system integrates diverse data sources including market data, on-chain metrics, and orderbook dynamics, translating these into unified buy/sell pressure signals. We demonstrate how machine learning models can effectively capture cross-timeframe relationships, enabling sub-second trading decisions with statistical confidence. The implementation addresses real-world challenges including data pipeline reliability, network latency, and operational risk management. Our results show consistent performance across various market conditions, with profit factor exceeding traditional approaches while maintaining manageable drawdown profiles.

\end{abstract}

\vspace{0.3cm}
\end{@twocolumnfalse}
]

\section{Introduction}

Cryptocurrency markets present unique opportunities for algorithmic trading due to their inherent volatility, 24/7 operation, and rich on-chain data availability. Unlike traditional financial markets, crypto markets provide unprecedented transparency through blockchain data, enabling sophisticated analysis of network activity, transaction flows, and market sentiment indicators.

The evolution of neural network architectures has revolutionized quantitative trading approaches, moving beyond traditional technical analysis toward deep learning models capable of processing multi-dimensional data streams in real-time \cite{lecun2015deep,sezer2020financial}. This paper describes a complete trading system that leverages these advances to achieve consistent profitability through systematic market exploitation.

Our approach recognizes that successful algorithmic trading requires more than accurate predictions—it demands robust data infrastructure, sophisticated risk management, and the ability to execute decisions within milliseconds of signal generation. The cryptocurrency market's unique characteristics, including extreme volatility and fragmented liquidity, necessitate specialized techniques for both signal generation and execution.

\section{Objective}

Machine learning transforms trading from intuition-based decision making to systematic pattern recognition across multiple data dimensions simultaneously \cite{krauss2017deep,heaton2017deep}. Statistical modeling provides the framework for quantifiable risk assessment and performance measurement, enabling traders to distinguish between skill and randomness in trading outcomes.

The primary objective is developing a comprehensive trading system that achieves four critical goals. First, gathering sufficient data sources containing genuine alpha signals through integration of market data, on-chain analytics, and sentiment information. Second, developing robust methods to understand and model underlying market trends across multiple timeframes. Third, training neural networks capable of predicting probable price direction with quantifiable confidence levels. Finally, executing these predictions on the smallest possible timeframes to maximize trading opportunities and minimize market exposure duration. This requires solving several interconnected problems: accurate trend identification across multiple timeframes, robust signal filtering to reduce false positives, and efficient execution in volatile market conditions.

Traditional approaches often fail because they rely on single-timeframe analysis or ignore the hierarchical nature of market movements. Our methodology addresses these limitations by explicitly modeling relationships between different temporal scales, from daily trends to sub-second orderbook dynamics. The underlying intuition is that buy/sell pressure manifesting in larger timeframes creates more likely conditions for orderbook dynamics to reflect this pressure through more granular and genuine order placement patterns. The system's success depends on its ability to synthesize information across these scales while maintaining computational efficiency required for high-frequency execution.

Statistical rigor underlies every component of the system, from hypothesis formulation through backtesting and live trading validation. This ensures that observed performance reflects genuine market inefficiencies rather than data mining artifacts or overfitting to historical patterns.

\section{Data Processing}

The data pipeline integrates three primary information sources into a unified framework for trading signal generation, utilizing historical data spanning from 2018 to present with the largest timeframe being daily intervals. Market data provides traditional price and volume information across multiple timeframes, forming the foundation for trend analysis and momentum indicators. On-chain data offers unique insights into network activity, transaction patterns, and holder behavior that can predict significant price movements or identify accumulation/distribution phases.

Orderbook data captures real-time supply and demand dynamics, enabling detection of short-term pressure imbalances that precede price movements \cite{hasbrouck1991measuring,cont2014price}. This microstructural information proves particularly valuable for execution timing and short-term direction prediction.

All data streams undergo careful normalization and synchronization to ensure consistent timing and scaling across different sources, with feature normalization performed using predetermined scaling parameters rather than dataset running statistics to prevent data snooping and ensure model stability.

For single asset, the system integrates multiple data sources including on-chain analytics capturing network activity and transaction patterns, traditional market data encompassing price and volume information, overall economic indicators such as S\&P 500 index movements, and sentiment data from GDELT datasets \cite{leetaru2013gdelt} providing 15-minute frequency sentiment data from global news and social media sources. This comprehensive data integration offers context for major market movements, particularly those driven by external events rather than technical factors \cite{tetlock2007giving,bollen2011twitter}. The multi-source approach helps the system adapt to regime changes and avoid trading during periods of elevated external uncertainty.

\begin{figure*}[t]
    \centering
    \includegraphics[width=\textwidth]{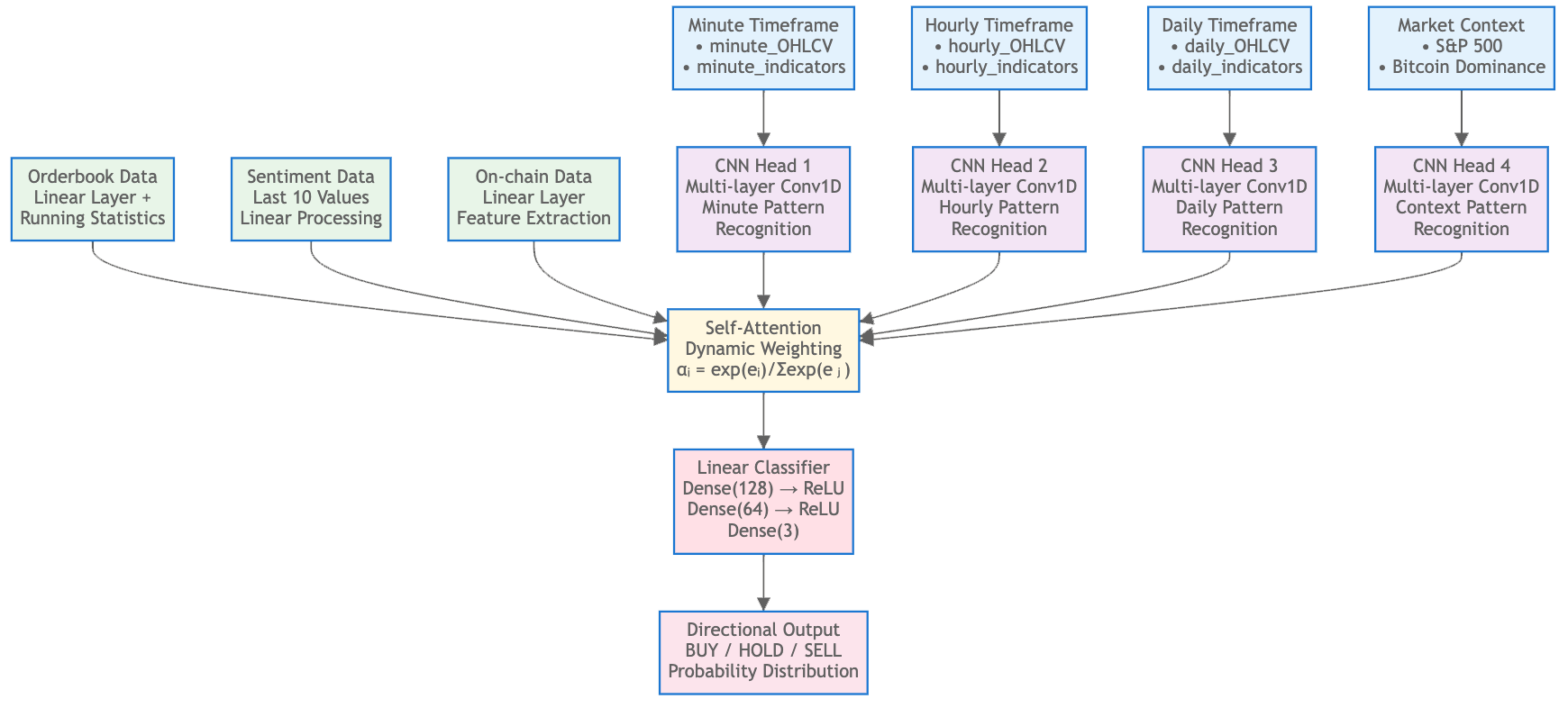}
    \caption{Direction Networks Architecture: Multi-head CNN processing of timeframe-specific market data (minute, hourly, and daily OHLCV with technical indicators) and broader market context (S\&P 500, Bitcoin dominance), combined with specialized linear processing of orderbook statistics, sentiment data, and on-chain metrics. Self-attention mechanism dynamically weights all feature inputs before final classification into directional trading signals.}
    \label{fig:direction_networks}
\end{figure*}

\section{Trend Networks}

The trend prediction component employs relatively simple neural networks designed for computational efficiency and interpretability. These networks process Bitcoin dominance metrics, network transaction volumes, and multi-timeframe moving averages to generate trend assessments on a continuous scale from -1 (strongly bearish) to +1 (strongly bullish).

The architecture prioritizes stability over complexity, using regularization techniques to prevent overfitting while maintaining sensitivity to genuine trend changes. Training data spans multiple market cycles to ensure robustness across different market regimes, including bull markets, bear markets, and sideways consolidation periods.

Bitcoin dominance serves as a primary input because of its strong correlation with overall cryptocurrency market sentiment. When Bitcoin dominance rises, altcoins typically underperform, while declining dominance often signals risk-on behavior across the broader crypto ecosystem.

Network transaction data provides insight into actual usage and adoption trends, which often precede price movements by days or weeks. The models learn to distinguish between genuine adoption signals and temporary spikes caused by network congestion or arbitrage activities.

Moving average analysis across daily, weekly, and monthly timeframes captures longer-term momentum while filtering out short-term noise. The neural networks learn optimal combinations of these timeframes rather than relying on predetermined technical analysis rules.

\section{Direction Networks}

Direction prediction utilizes a sophisticated multi-head CNN architecture where each head specializes in processing different temporal patterns for performance optimization \cite{borovykh2017conditional,tsantekidis2017forecasting}. The system employs three main CNN heads organized by timeframe: minute-scale, hourly-scale, and daily-scale processing. Each head processes OHLCV data, selected technical indicators to minimize noise, sentiment data, and on-chain metrics. Additionally, the hourly and daily heads incorporate broader market context including S\&P 500 index and Bitcoin dominance metrics for enhanced pattern recognition across multiple scales.

Each CNN head consists of multiple convolutional layers designed to extract temporal patterns specific to its assigned timeframe. The minute-focused head processes high-frequency signals for immediate directional changes, while hourly and daily heads capture medium to long-term trend patterns and incorporate S\&P 500 and Bitcoin dominance metrics to understand broader market regime influences.

Complementing the CNN processing, orderbook data undergoes specialized linear processing combined with running statistical analysis including metrics such as biggest order size and gap measurements in the last minute and hour to capture microstructural supply-demand imbalances. Sentiment data from GDELT's 15-minute global news feed, represented as the last ten values, feeds into each CNN head (daily, hourly, and minute-scale) for contextual weighting of market mood and external event impacts. On-chain data such as transaction count and volume metrics are integrated into the CNN heads alongside other technical indicators for comprehensive pattern recognition.

The soft attention mechanism serves as a dynamic selection network, adaptively weighting outputs from the three CNN heads and the orderbook linear processor based on current market conditions \cite{vaswani2017attention,qin2017dual}. The intuition behind this approach is to select features that are relevant, ideally allowing each head to contribute with certain weight based on the current market situation. Given feature representations from different processing heads, the attention weights are computed as:

\begin{align}
e_i &= f_{att}(x_i, h) \\
\alpha_i &= \frac{\exp(e_i)}{\sum_{j=1}^{n} \exp(e_j)} \\
c &= \sum_{i=1}^{n} \alpha_i x_i
\end{align}

where $f_{att}$ represents the learned attention function, $h$ captures current market context, $\alpha_i$ are normalized attention weights, and $c$ is the context-weighted representation. During volatile periods, the system increases weight on orderbook and minute-scale features, while trending markets emphasize daily patterns and sentiment indicators.

The network training employs multi-class cross-entropy loss with L2 regularization to prevent overfitting and improve generalization across the three directional classes: buy, sell, and hold.

The final classification layers process the attention-weighted features through dense neural networks to produce categorical directional predictions. This architecture enables specialized pattern recognition across multiple data types while maintaining computational efficiency required for real-time trading applications. The direction networks are relatively compact with approximately 520,000 parameters, optimizing for both performance and inference speed. Direction networks undergo periodic retraining as market regime shifts occur, ensuring model relevance and preventing performance degradation over time.

\begin{table*}[t]
\centering
\caption{Network Architecture Performance Comparison}
\label{tab:architecture_comparison}
\begin{tabular}{lcc}
\toprule
\textbf{Architecture} & \textbf{Profit Factor} & \textbf{Notes} \\
\midrule
CNN + Soft Attention (Current) & 1.15 & Optimal performance \\
LSTM (instead of CNN) & 1.08 & Slower inference \\
CNN without Attention & 0.98 & Very low confidence scores \\
LSTM without Attention & 0.95 & Poor performance \& slow \\
\bottomrule
\end{tabular}
\end{table*}

The architecture comparison demonstrates the critical importance of the soft attention mechanism, with the CNN-only variant producing very low confidence scores despite similar directional accuracy. The LSTM-based approach, while functional, showed inferior performance and significantly slower inference times compared to the optimized CNN architecture.

\section{Inference}

The inference process dynamically selects appropriate direction networks based on current trend conditions identified by the trend networks. This two-stage approach ensures that directional predictions come from models specifically trained for current market conditions, improving accuracy and confidence.

\begin{algorithm}
\caption{Real-time Inference Pipeline}
\begin{algorithmic}[1]
\Procedure{InferenceStep}{$market\_data$, $trend\_networks$, $direction\_networks$}
    \State $trend\_score \gets$ EvaluateTrend($market\_data$, $trend\_networks$)
    \State $selected\_networks \gets$ SelectNetworks($trend\_score$)
    \State $predictions \gets []$
    \State $confidences \gets []$
    \For{each $network$ in $selected\_networks$}
        \State $pred, conf \gets network$.Predict($market\_data$)
        \State $predictions$.append($pred$)
        \State $confidences$.append($conf$)
    \EndFor
    \State $ensemble\_pred \gets$ WeightedVote($predictions$, $confidences$)
    \State $consensus\_score \gets$ CalculateConsensus($predictions$)
    \State $final\_confidence \gets$ CombineScores($confidences$, $consensus\_score$)
    \If{$final\_confidence > threshold_{high}$}
        \State \Return ExecuteTrade($ensemble\_pred$, position\_size = $large$)
    \ElsIf{$final\_confidence > threshold_{low}$}
        \State \Return ExecuteTrade($ensemble\_pred$, position\_size = $small$)
    \Else
        \State \Return NoAction()
    \EndIf
\EndProcedure
\end{algorithmic}
\end{algorithm}

Multiple direction networks evaluate each trading opportunity, generating an ensemble prediction with associated confidence scores. High-confidence unanimous decisions trigger immediate trading actions, while lower-confidence or conflicting signals may result in position sizing adjustments or trade abstention.

The confidence scoring mechanism incorporates both individual network certainty and inter-network agreement. High individual certainty with strong consensus indicates ideal trading conditions, while high certainty with poor consensus suggests regime uncertainty requiring more conservative position sizing.

Real-time evaluation occurs continuously as new data arrives, enabling the system to adapt quickly to changing market conditions. The inference pipeline achieves sub-50ms latency per prediction, with complete trade execution typically requiring 100-300ms including network latency and exchange processing time.

\section{Hypothesis}

Our core hypothesis rests on the hierarchical nature of market information flow across different timeframes. We posit that larger timeframes contain information that constrains and influences shorter timeframe movements, while microstructural data provides early signals of direction changes that propagate upward through the timeframe hierarchy.

Daily trends establish the primary market direction, creating a bias that influences hourly patterns despite higher-frequency noise. This relationship enables models trained on longer timeframes to provide valuable context for shorter-term predictions, improving accuracy and reducing false signals.

Conversely, orderbook pressure imbalances often precede visible price changes, providing early warning signals that can be detected in real-time. These microstructural patterns tend to cascade upward, influencing minute-by-minute price movements and eventually affecting longer-term trends.

The key insight is that information flows bidirectionally across timeframes. While longer timeframes provide directional bias, shorter timeframes reveal the timing and magnitude of moves. Optimal trading systems must capture both dimensions simultaneously rather than analyzing them in isolation.

This theoretical framework enables the system to maintain statistical edges across different market conditions. During strong trends, longer-timeframe signals dominate, while in volatile or ranging markets, microstructural signals become more important for timing entries and exits.

\section{Trading}

The low parameter count of direction networks enables parallel inference across multiple trading opportunities simultaneously. Combined with pre-computed trend assessments, the system achieves sub-second decision-making capabilities essential for high-frequency trading approaches. Predictions are made at the tick level, utilizing the smallest possible timeframes to capture immediate market movements and maximize the number of independent trading opportunities.

Trading execution occurs at the tick level, with the complete decision-to-execution cycle typically requiring 100-300ms including inference time, network latency, and exchange processing. Each trade follows a rapid cycle: identify opportunity, execute entry, monitor position, and execute exit based on either profit targets or stop conditions. This approach maximizes the number of independent trading opportunities while minimizing exposure duration.

Position sizing adapts dynamically based on confidence scores and current market volatility. Higher-confidence signals with favorable trend alignment receive larger allocations, while uncertain or counter-trend opportunities receive minimal position sizes or may be skipped entirely.

The system maintains strict risk controls throughout the execution process, including maximum position limits, correlation-based exposure limits, and real-time drawdown monitoring. These controls operate independently of the signal generation process, providing multiple layers of protection against model failures or market disruptions.

\section{Testing}

Backtesting methodology emphasizes statistical rigor over optimistic projections, using out-of-sample testing to validate model performance. The testing framework simulates realistic trading conditions including 0.05\% transaction costs and accounts for variance in slippage to ensure results reflect achievable performance.

Metrics measured include maximum profit factor, sharpe, drawdown, win rate to provide comprehensive performance assessment.

Multiple market regimes receive equal attention during testing to ensure robustness across different conditions. This includes bull markets, bear markets, high-volatility periods, and low-volatility consolidation phases that challenge different aspects of the trading system.

The architecture comparison results presented in Table 1 demonstrate the superior performance of the proposed CNN with soft attention approach, achieving a profit factor of 1.15 compared to alternative configurations. The LSTM-based approaches showed consistently lower performance, with LSTM without attention producing the worst results at 0.95 profit factor. Notably, removing the attention mechanism from the CNN architecture resulted in very low confidence scores despite maintaining reasonable directional accuracy, highlighting the critical role of attention in reliable signal generation.

\section{Production Challenges}

Production deployment revealed significant operational challenges beyond statistical modeling concerns. Websocket connection instability created data gaps that could trigger false signals or prevent proper risk management. Orderbook lag during high-volatility periods introduced execution delays that eroded theoretical profits.

Missing dataframes due to network issues or exchange downtime created the highest risk scenarios, as the system could make decisions based on incomplete information. Implementing robust data validation and fallback mechanisms proved more critical than optimizing prediction accuracy.

Latency issues compound rapidly in high-frequency trading environments, where millisecond delays can eliminate profitable opportunities. Network topology, server co-location, and optimized code paths became as important as model performance for overall system success.

These operational realities highlight the gap between backtesting environments and live trading conditions. While statistical models provide the foundation for profitable trading, robust engineering and operational procedures determine whether theoretical advantages translate into actual profits.

\section{Conclusion}

This paper demonstrates that neural network-based approaches can achieve consistent profitability in cryptocurrency markets through systematic exploitation of cross-timeframe relationships and microstructural patterns. The multi-stage architecture successfully combines trend analysis with high-frequency direction prediction, enabling robust performance across diverse market conditions.

The integration of multiple data sources into unified pressure indicators proves effective for capturing market dynamics that individual indicators might miss. The hierarchical timeframe approach provides both theoretical foundation and practical advantages for real-time trading systems.

\bibliographystyle{plainnat}
\bibliography{references}

\end{document}